\documentclass[twocolumn,prl,superscriptaddress,showpacs]{revtex4}
\usepackage{epsf,graphicx,amssymb,subfig,color}
\begin{document}
\title{Dipole-Quadrupole dynamics during magnetic field reversals}

\author{Christophe Gissinger} \affiliation{Department of Astrophysical
  Sciences, Princeton University, Princeton, NJ 08544.}

\def\bfnabla{\mbox{\boldmath $\nabla$}}

\begin{abstract}
The shape and the dynamics of reversals of the magnetic field in a
turbulent dynamo experiment are investigated. We report the evolution
of the dipolar and the quadrupolar parts of the magnetic field in the
VKS experiment, and show that the experimental results are in good
agreement with the predictions of a recent model of reversals: when
the dipole reverses, part of the magnetic energy is transferred to the
quadrupole, reversals begin with a slow decay of the dipole and are
followed by a fast recovery, together with an overshoot of the
dipole. Random reversals are observed at the borderline between 
stationary and oscillatory dynamos.
\end{abstract}
\maketitle 

Despite a large variability of the internal structure of planets and
stars, most of the observed astrophysical bodies possess a coherent
large scale magnetic field. It is widely accepted that these natural
magnetic fields are self-sustained by dynamo action \cite{Dormy}.
Although reversals of the magnetic field in planetary and stellar
dynamos are now considered to be a common feature, they still remain
poorly understood. Whereas the Sun shows periodic oscillations of its
magnetic field, the polarity of the Earth's dipole field reverses
randomly. During the last decades, several mechanisms have been
proposed for geomagnetic reversals, among which we can mention, the
analogy with a bistable oscillator \cite{Hoyng01}, a mean-field dynamo
model \cite{Stefani05}, or interaction between dipolar and higher
axisymmetric components of the magnetic field
\cite{McFadden91,Clement04}. The comprehension of
  dynamo reversals have also benefited from direct numerical
  simulations of the MHD equations, which have displayed several
  possible mechanisms for reversals \cite{Glatz95}, \cite{CoeGlatz06},
  \cite{Gissinger10}.

Reversals of a dipolar magnetic field have also been reported in the
VKS (Von Karman Sodium) dynamo experiment \cite{Berhanu07}. In this
experiment, periodic or chaotic flips of polarity can be observed
depending on the magnetic Reynolds number. Based on these results, a
model for reversals has been recently proposed by P\'etr\'elis and
Fauve~\cite{Petrelis08}. It relies on the interaction between the
dipolar and the quadrupolar magnetic components, and describe
transitions to periodic oscillations or randomly reversing dynamos. It
has been claimed that such a mechanism could apply to the reversals of
the Earth magnetic field \cite{Petrelis09}, and temptatively be
connected to the periodic oscillations of the Solar
dynamo. Unfortunately, as for many other models, the lack of
observations of the magnetic field during a reversal limits a direct
comparison with the actual geomagnetic reversals. From this point of
view, the VKS experiment is a unique opportunity to test the validity
of different models of turbulent reversing dynamos. In particular, the
model~\cite{Petrelis08} makes predictions about the dynamics that are
easily confrontable to experimental results.  We propose a simple way
to analyze data from the VKS experiment in order to test this
model. To wit, we extract from the data the dipolar and the
quadrupolar components of the magnetic field.  We show that the
characteristics of the reversals in the VKS experiment are in very
good agreement with the predictions, and that the dynamics of the
magnetic field in this turbulent dynamo mainly result from an
interaction between dipolar and quadrupolar modes.

\begin{figure}
\centerline{
\epsfysize=52mm 
\epsffile{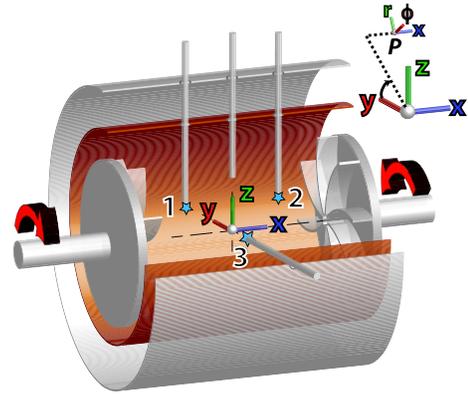} }
\caption{(Color online) Sketch of the VKS experiment}
\label{vks}
\end{figure}

In the VKS experiment, a turbulent von Karman flow of liquid sodium is
generated inside a cylinder by two counter-rotating impellers, with
independent rotation frequencies $F_1$ and $F_2$ (see figure
\ref{vks}, and~\cite{Berhanu09} for the description of the
set-up). When $F_1= F_2$, the system is invariant about a rotation of
an angle $\pi$ around any axis located in the mid-plane. On the
contrary, if the impellers rotate at different rates, this symmetry,
hereafter referred to as the $R_\pi$ symmetry, is broken. The dynamics
of the magnetic field observed in the experiment strongly depend on
this symmetry. When $F_1=F_2$, a statistically stationary magnetic
field with either polarity is generated, with a dominant axial dipolar
component. Dynamical regimes, including periodic oscillations and
chaotic reversals of the magnetic field, are observed only when the
$R_\pi$ symmetry is broken ($F_1\ne F_2$).

Previous studies have suggested that the evolution of the magnetic
field in the VKS experiment results from low dimensional dynamics,
involving only a few modes in interaction~\cite{Ravelet08}. This can
be ascribed to the proximity of the bifurcation threshold and to the
smallness of the magnetic Prandtl number in liquid metals
($Pm<10^{-5}$). Indeed, in the low-$Pm$ regime, the magnetic field is
strongly damped compared to the velocity field. Hence, the dynamics
are governed only by a small number of magnetic modes. Based on this
observation, P\'etr\'elis and Fauve \cite{Petrelis08} have proposed
that close to the dynamo threshold, the magnetic field can be
decomposed into two components:
\begin{equation}
{\bf B}=D(t) \, {\bf d}({\bf r}) + Q(t) \, {\bf q}({\bf r})
\end{equation}
where $D$ (respectively $Q$) represents the amplitude of dipolar ${\bf
  d}({\bf r})$ (resp. quadrupolar ${\bf q}({\bf r}$)) component of the
field.  As emphasized in \cite{Petrelis08}, these components do not
only involve a dipole or a quadrupole, but also all the higher
components with the same symmetry in the transformation $R_\pi$. In
other words, ${\bf d}({\bf r})$ (resp. ${\bf q}({\bf r})$) is the
antisymmetric (resp. symmetric) part of the magnetic field.

The evolution equations for $D(t)$ and $Q(t)$ can then be obtained by
symmetry arguments (see~\cite{Petrelis08} for a detailed description
of the model). 
Since ${\bf d} \rightarrow - {\bf d}$ and ${\bf q} \rightarrow {\bf
  q}$ in the transformation $R_\pi$, dipolar and quadrupolar modes
cannot be linearly coupled when $F_1=F_2$. Breaking the $R_\pi$
symmetry by rotating the impellers at different speeds allows a linear
coupling between dipolar and quadrupolar modes. For a sufficiently
strong symmetry-breaking, this coupling can generate a limit cycle
that involves an energy transfer between dipolar and quadrupolar
modes.  This mechanism has been recently validated on a numerical
model of the VKS experiment~\cite{Gissinger09} and also in the case of
a mean-field $\alpha^2$ dynamo model \cite{Gallet09}.

Two scenarios of transition from a stationary dynamo to a periodically
reversing magnetic field can be described in the framework of this
dipole-quadrupole model. When the coupling is such that the system is
close to both a stationary and a Hopf bifurcation, i.e. in the
vicinity of a codimension-two bifurcation point, one can have
bistability between a stationary and a time periodic reversing
dynamo~\cite{Berhanu09}. We thus get a subcritical transition from a
stationary dynamo to a periodic one with a finite frequency at
onset. Turbulent fluctuations can generate random transitions between
these two regimes~\cite{Berhanu10}.  Far from this codimension-two
point, a reversing magnetic field can be generated through an Andronov
bifurcation when the stationary state disappears through a saddle-node
bifurcation~\cite{Petrelis08}. Then, the frequency of the limit cycle
vanishes at onset. In the vicinity of this transition, turbulent
fluctuations drive random reversals of the magnetic field. As a
consequence, random reversals always occur at the borderline between
stationary and oscillatory dynamos. This simple mechanism also yields
several predictions about the shape of the reversals. First, when the
dipole $D$ vanishes, part of the magnetic energy is transferred to the
quadrupole $Q$. An overshoot of the dipolar amplitude is expected
after each reversal. Random reversals are asymmetric. During a first
phase, fluctuations push the system from the stable solution to the
unstable one, thus acting against the deterministic dynamics. This
phase is slow compared to the one beyond the unstable fixed point,
where the system is driven to the opposite polarity under the action
of the deterministic dynamics.\\

\begin{figure}
\vskip -8mm
\centerline{
\epsfysize=80mm 
\epsffile{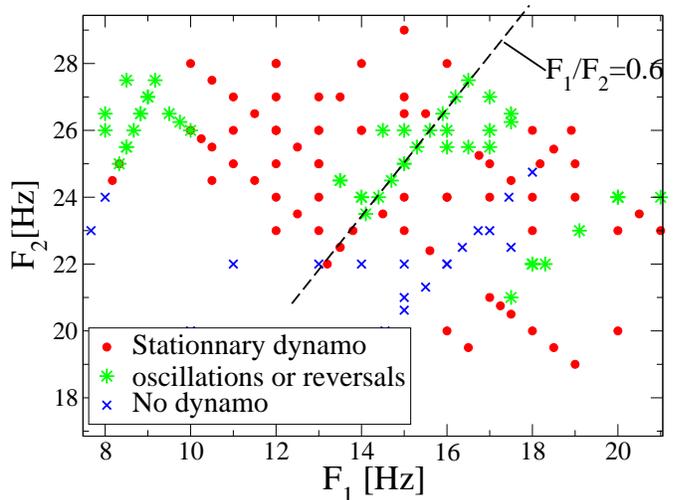} }
\vskip -3mm
\caption{(Color online) Parameter space: (cross) no dynamo, (circle) stationary dynamos,
  (star) oscillatory or random reversing dynamos.}
\label{param}
\end{figure}

\begin{figure}
\vskip -2mm
\centerline{
\epsfxsize=70mm 
\epsffile{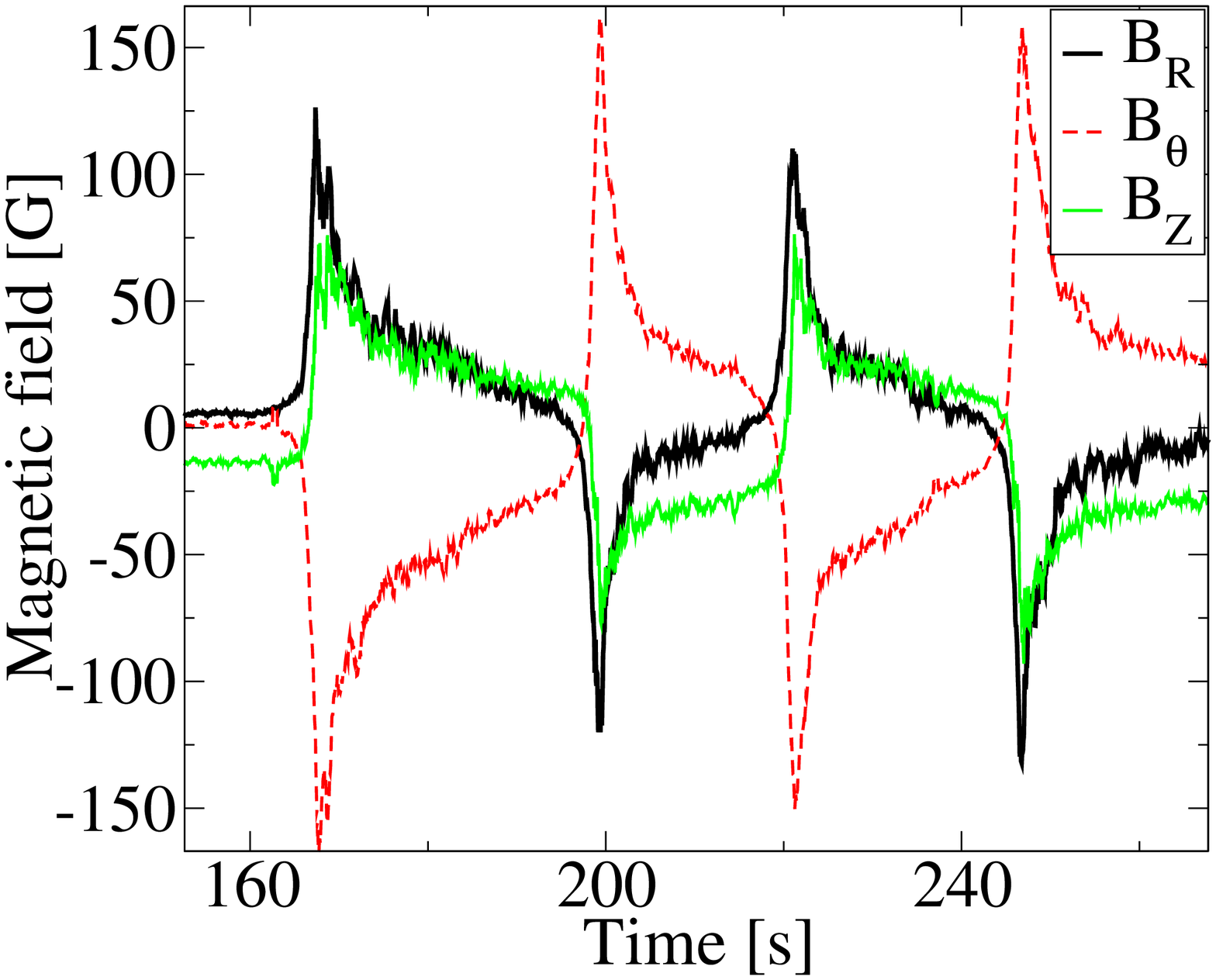} }
\centerline{
\epsfxsize=70mm 
\epsffile{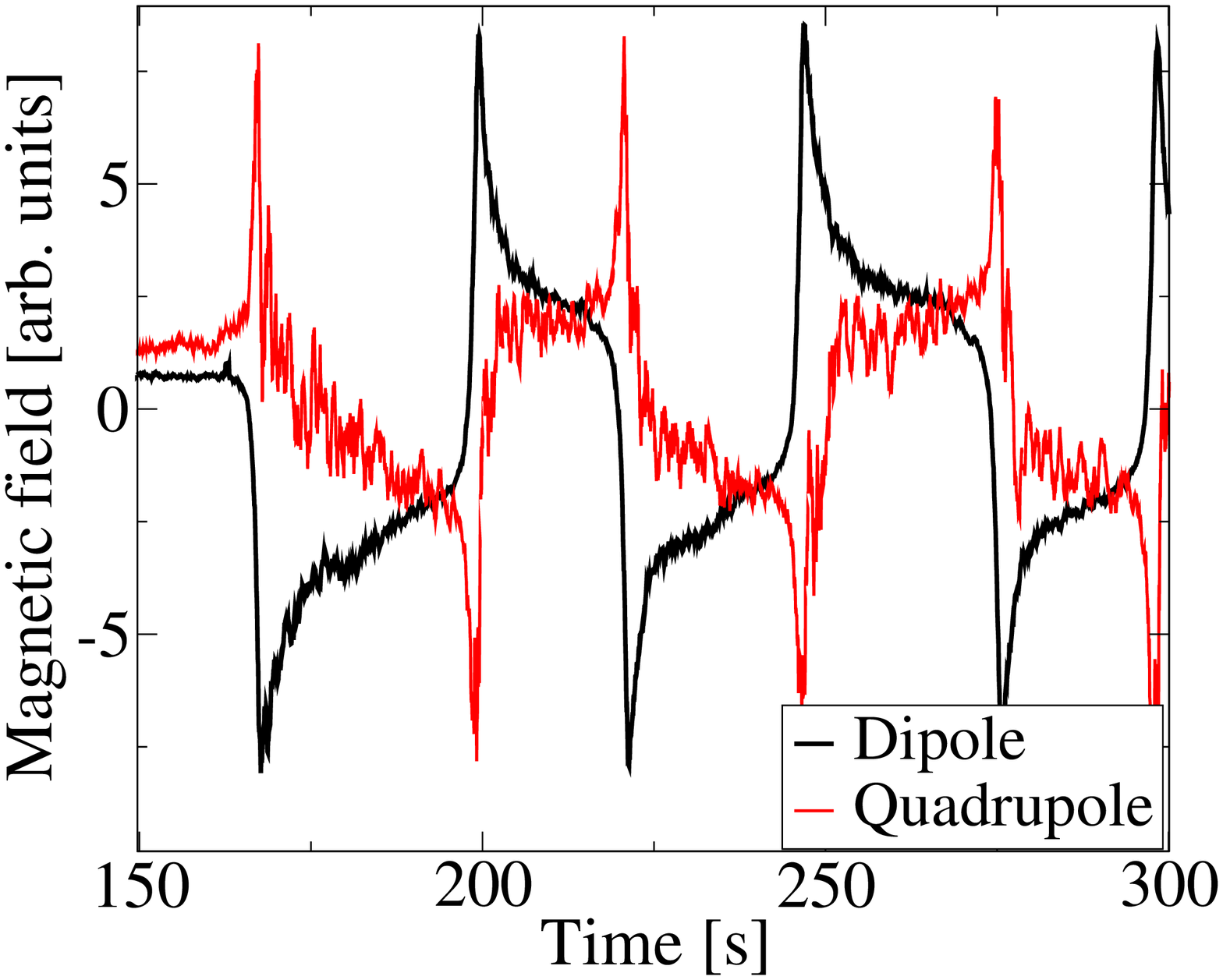} }
\caption{(Color online) Top: time recordings of the three components of
    the magnetic field. The rotation frequencies are increased from
    $14.4/24$ Hz to $15/25$ Hz, leading to a transition from a
    stationary low amplitude dynamo regime to a limit cycle. Bottom:
    Behavior of the dipolar and quadrupolar parts of the magnetic
    field. Note the transfer between the two components during
    reversals.}
\label{periodic}
\end{figure}

In this paper, we use data of the VKS experiment in order to
reconstruct the dipolar and the quadrupolar parts of the magnetic
field and study their behavior in the time dependent
regimes. Time-dependent regimes only occur for specific values of
$F_1$ and $F_2$, inside three delimited regions of the parameter
space~\cite{Berhanu10}. We will focus on the regimes observed when
following the line $F_1/F_2=0.6$ in the parameter space. Figure
\ref{param} shows that when the rotation rates are increased along
this line, one first bifurcates to a stationary dynamo, then to
time-dependent regimes.
Figure \ref{periodic}(top) shows the time-recordings of the three
components of the magnetic field close to the fastest disk, displaying
the bifurcation from stationary to time-dependent dynamo when the
frequencies of the two disks are increased from $14.4/24$ Hz
($F_1+F_2=38.4$ Hz) to $15/25$ Hz ($F_1+F_2=40$ Hz). After a short
transient state, the three components of the magnetic field undergo a
transition to nearly periodic oscillations.

However, it is hard to test the pertinence of the
model~\cite{Petrelis08} from the time recording of the magnetic field
at a single point. Using measurements obtained from two probes $1$ and
$2$, symmetric with respect to the mid-plane, we compute the dipolar
part, $D(t) d_i=(B_i(1,t)+B_i(2, t))/2$ and the quadrupolar part,
$Q(t) q_i=(B_i(1, t)-B_i(2,t))/2$. In order to obtain observables
which are independent of the spatial component $i$, each of these
vectors is projected on its value at a given time $t_0$. We thus
extract $D(t)$ and $Q(t)$ up to a multiplicative constant. In the
measurements displayed here, the dipolar and quadrupolar components
are projected on their stationary values obtained at $F_1+F_2=38.4$
Hz.  Note however that different methods could be used to reconstruct
these amplitudes. In particular, plotting the sum and the difference
of a given component does not change the qualitative
behavior~\cite{GissingerThesis}. Figure \ref{periodic}(bottom) shows
the evolution of $D(t)$ and $Q(t)$ during periodic oscillations of the
magnetic field. We observe that when the dipole vanishes, the
amplitude of the quadrupole reaches its maximum. This shows that the
field reversals observed in the VKS experiment do not correspond to a
vanishing magnetic field, but rather to a change of shape from a
dominant dipolar field to a quadrupolar one. Immediately after each
reversal, one can also note that during its recovery, the dipolar
amplitude strongly overshoots its mean value. Therefore, in agreement
with the model~\cite{Petrelis08}, reversals in the VKS experiment
involve a strong competition between dipolar and quadrupolar
components of the magnetic field.

\begin{figure*}
\vskip -2mm
\centerline{
\epsfysize=48mm 
\subfloat[]{\label{cyclea}}\epsffile{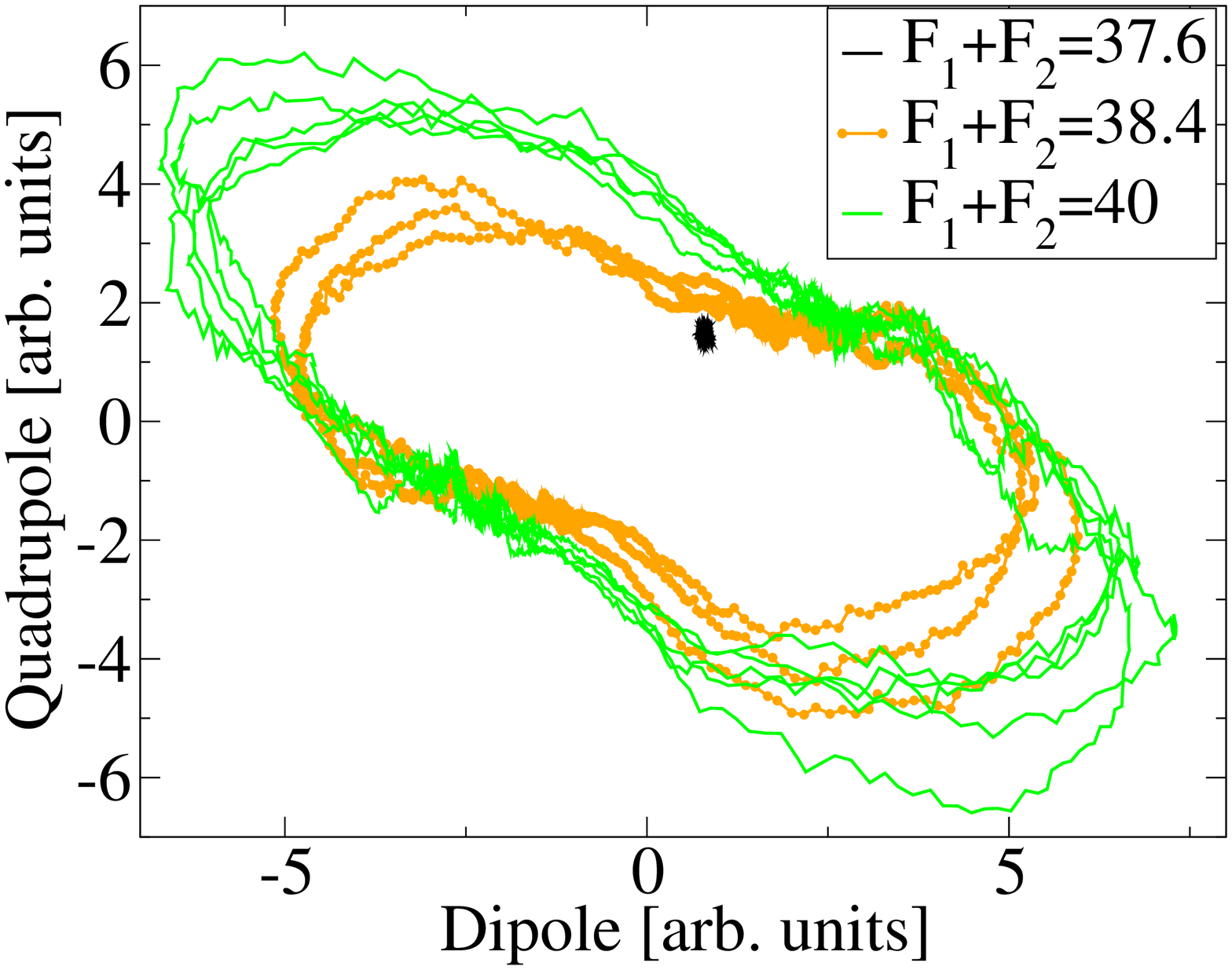}
\epsfysize=48mm 
\subfloat[]{\label{cycleb}}\epsffile{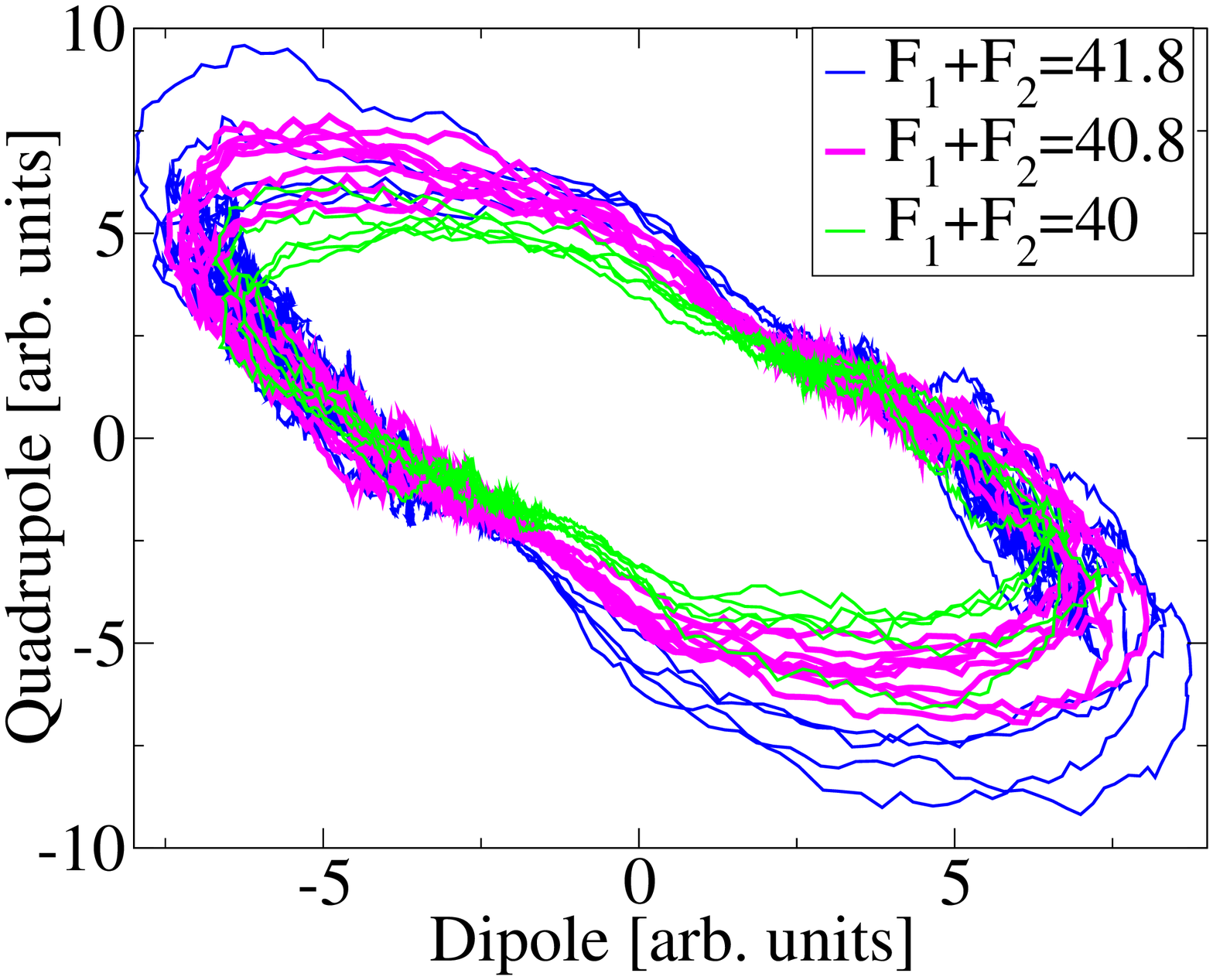}
\epsfysize=48mm 
\subfloat[]{\label{cyclec}}\epsffile{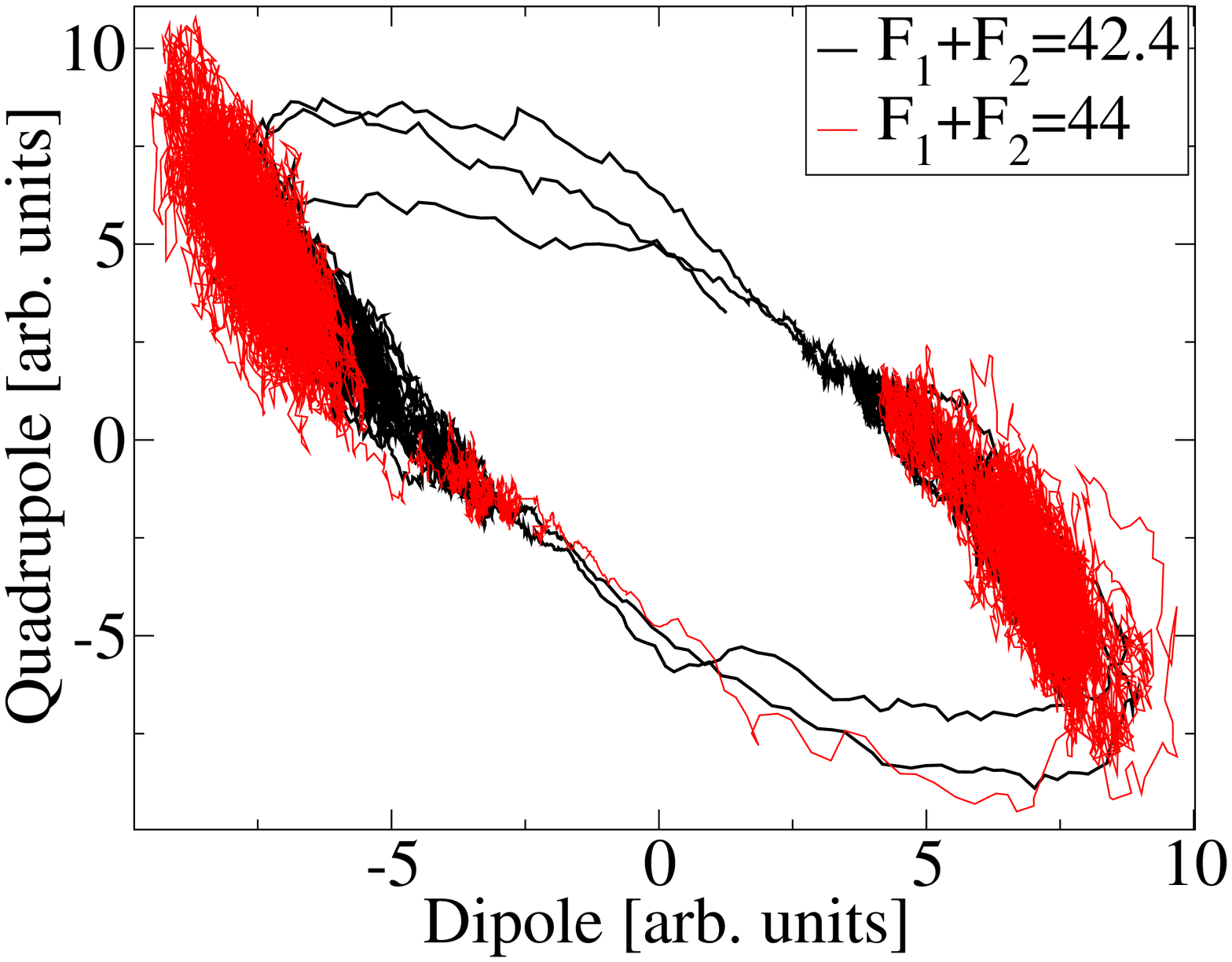}
}
\vskip -2mm
\caption{(Color online) (a): Evolution of the magnetic field in the
    phase space $(D,Q)$ at low frequencies: there exists a range of
    bistability in which a stationary dynamo (black dot) and a
    periodic limit cycle (orange circles) are both metastable. (b): Evolution
    of the limit cycle when $F_1+F_2$ is increased from $40$ to $41.8$
    Hz. (c): Chaotic reversals obtained for large values of $F_1+F_2$
    in the vicinity of a saddle-node bifurcation.}
\label{cycle}
\end{figure*}

The decomposition between dipolar and quadrupolar components is not
only relevant to study these oscillations but is also useful to follow
the bifurcations observed along the line $F_1/F_2 = 0.6$ in
figure~\ref{param}. We now investigate the evolution of the dynamics
in the phase space $(D, Q)$ displayed in figure \ref{cycle} as
$F_1+F_2$ is modified.  The limit cycle described in
figure~\ref{periodic} bifurcates from a low amplitude stationary
dynamo when $F_1+F_2$ is increased from $38.4$ to $40$ Hz.  This limit
cycle is shown in green in figure~\ref{cyclea}. When $F_1+F_2$ is
decreased again to $38.4$ Hz, a smaller amplitude limit cycle is
obtained (orange curve, circles) instead of a fixed point. We need to
decrease the rotation frequencies further to recover the low amplitude
stationary dynamo (black dot). Therefore, this transition is
hysteretic and within some frequency range we have bistability
involving stationary and time-periodic dynamos. This oscillation
appears at finite amplitude and finite period~\cite{Berhanu10}. The
oscillation of figure \ref{periodic} displays a slowing down in the
vicinity of two symmetric fixed points, as expected for a system close
to the saddle-node bifurcation of Andronov type. Note however that the
onset of the cycle when $F_1+F_2$ is increased, does not correspond to
such a saddle-node bifurcation, since these two stagnation points are
distinct from the low amplitude stationary dynamo regime obtained at
$F1 + F2 = 38.4$. In fact, this transition from a low amplitude
stationary magnetic field to an oscillatory regime at finite period,
rather corresponds to the model taken close to its codimension-two
bifurcation point \cite{Berhanu09}.

When $F_1+F_2$ is increased further, the amplitude of the limit cycle
continuously increases (figure~\ref{cycleb}). In addition, the system
slows down in the vicinity of two points $(\pm D_s, \pm Q_s)$
(figure~\ref{cyclec}). Thus, the period of the limit cycle
significantly increases. For $F_1+F_2 = 44$ Hz, the systems stops on
one of these two fixed points and we get a stationary dynamo (although
we cannot rule out the occurrence of other reversals with a longer
experiment). As explained in the framework of the
model~\cite{Petrelis08}, this second transition corresponds to a
saddle-node bifurcation or more precisely an Andronov bifurcation: the
stable fixed points $(\pm D_s, \pm Q_s)$ collide with unstable fixed
points $(\pm D_u, \pm Q_u)$ when $F_1+F_2$ is decreased and
disappear. A limit cycle is thus created, and its period is expected
to diverge in the vicinity of the saddle-node bifurcation.
Turbulent fluctuations of course saturate this divergence by kicking
the system away from the points $(\pm D_s, \pm Q_s)$ where it slows
down. They also strongly modify the dynamics on the other side of the
bifurcation. Indeed, when the stable and unstable fixed points are
very close one to the other, turbulent fluctuations can randomly drive
the system from a stable fixed point to its neighboring unstable one,
and thus trigger a reversal of the magnetic field.  Therefore, random
reversals are expected in the vicinity of the saddle-node
bifurcation~\cite{Petrelis08}. This is what is observed here as shown
below.

\begin{figure}
\centerline{
\hskip -6mm
\epsfxsize=80mm 
\epsfysize=60mm 
\epsffile{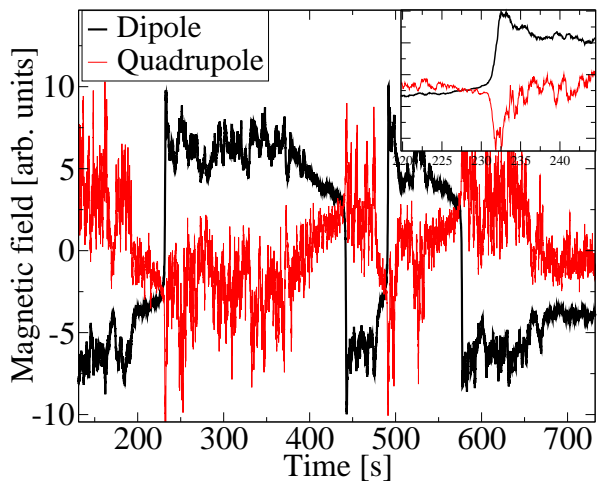} }
\caption{(Color online) Time evolution of the magnetic field during a
    regime of chaotic reversals, for $F_1+F_2=42.4$ Hz. Inset: zoom on
    the reversal occuring at $t=231$s.}
\label{timeREV}
\end{figure}
 
Figure \ref{timeREV} displays the time-recordings of the dipolar and
quadrupolar components for $F_1+F_2=42.4$ Hz. We observe that both
components fluctuate around constant values as if they have reached a
stable fixed point.  The time spent in both polarities is random but
much longer than the magnetic diffusion time scale (of order $1$
s). One also clearly observes that the amplitude of the dipole slowly
decreases before rapidly changing sign. In the phase space $(D, Q)$
displayed in figure~\ref{cyclec}, this slow decay corresponds to
random motion in the regions in the form of elongated spots located
along the limit cycle in the vicinity of the fixed points. Indeed, the
motion from each stable fixed point to the neighboring unstable one,
occurs under the influence of fluctuations acting against the
deterministic dynamics. It is thus a slow random drift compared to the
fast reversal phase driven by the deterministic dynamics once the
system has been pushed beyond the unstable fixed
point. Figure~\ref{cyclec} also show that the spots become more and
more elongated when $F_1+F_2$ is increased. This tells that the
distance between each stable fixed point and its unstable neighbor
increases.  Correspondingly, reversals are less frequent. For
$F_1+F_2=44$ Hz (red (light grey) cycle in figure~\ref{cyclec}),
fluctuations can hardly drive reversals. As for periodic oscillations
obtained above the Andronov bifurcation, the modal decomposition $(D,
Q)$ underlines the short transfer from an axial dipole to a
quadrupolar magnetic field during random reversals obtained below the
bifurcation threshold (see inset of figure \ref{timeREV}). The dipolar
amplitude displays the expected behavior, characterized by a slow
decay followed by a rapid recovery, and showing a typical overshoot
after each reversal.  Evolution in the phase space $(D,Q)$ also
illustrate how the transfer between dipolar and quadrupolar components
yields very robust cycles, systematically avoiding the origin $B=0$.\\

In conclusion, we have used a simple method to extract the dipolar
(antisymmetric) and quadrupolar (symmetric) components of the magnetic
field in the VKS experiment. We have shown that this decomposition
allows to investigate the morphology of the magnetic field during
reversals, and to compare experimental results to the predictions of a
recent model proposed in \cite{Petrelis08}. We have shown that the
results of the VKS experiment are in very good agreement with these
predictions:

- reversals are characterized by a strong transfer to the quadrupole
when the dipole vanishes,

- the dipolar mode systematically displays an overshoot after each reversal,

- random reversals are asymmetric, i.e. involve two phases: a slow one
triggered by turbulent fluctuations followed by a fast one mostly
governed by the deterministic dynamics.

This agreement between the VKS experiment and the model has
significant consequences. It first shows that a fluid dynamo, even
generated by a strongly turbulent flow, can exhibit low dimensional
dynamics, involving mostly dipolar and quadrupolar modes. Furthermore,
because such a model is based on symmetry arguments, the mechanisms
described here are expected to apply beyond the VKS experiment. For
instance, although 3-dimensional simulations do not involve a similar
level of turbulence, a transfer between dipole and quadrupole during
reversals has been observed in several numerical studies of the
geodynamo \cite{Glatz95}, \cite{CoeGlatz06}. This is consistent with
indirect evidences from paleomagnetic measurements, suggesting a
dipole-quadrupole interaction~\cite{McFadden91} and asymmetric
reversals~\cite{Valet05}. Observations of the Sun's magnetic field also
suggest a transfer between dipolar and higher components
\cite{Knaack05}. In numerical simulations based on the VKS experiment
\cite{Gissinger10}, a good agreement with the three predictions
reported here has been obtained, but only when the magnetic Prandtl
number is sufficiently small. In this context, our simple method could
be used to investigate data from numerical simulations of the
geodynamo at low magnetic Prantl number. This opens new perspectives
to understand the dynamics of planetary and stellar magnetic fields
with a simple and low dimensional description.

\begin{acknowledgments}
I aknowledge my colleagues of the VKS team with whom the experimental
data used here have been obtained~\cite{Berhanu10} and
ANR-08-BLAN-0039-02 for support. I also thank anonymous referees for
very helpful comments.
\end{acknowledgments}


\end{document}